\documentclass[aps,prb,twocolumn,showpacs,superscriptaddress,floatfix]{revtex4-1}  
\usepackage[english]{babel}
\usepackage[utf8]{inputenc}
\usepackage{SIunits}
\usepackage{graphicx}
\usepackage{dcolumn}
\usepackage{amsmath}
\usepackage{natbib}
\usepackage{bm}

\begin{document}

\title{Polarization resolved magneto-Raman scattering of graphene-like domains on natural graphite}
\author{M.~K\"{u}hne}
\affiliation{LNCMI-CNRS (UJF, UPS, INSA), BP 166, 38042 Grenoble
Cedex 9, France}
\author{C.~Faugeras}\email{clement.faugeras@lncmi.cnrs.fr}
\affiliation{LNCMI-CNRS (UJF, UPS, INSA), BP 166, 38042 Grenoble
Cedex 9, France}
\author{P.~Kossacki}
\affiliation{LNCMI-CNRS (UJF, UPS, INSA), BP 166, 38042 Grenoble
Cedex 9, France} \affiliation{Institute of Experimental Physics,
University of Warsaw, Hoza 69, 00-681 Warsaw, Poland}
\author{A.~A.~L.~Nicolet}
\affiliation{LNCMI-CNRS (UJF, UPS, INSA), BP 166, 38042 Grenoble
Cedex 9, France}
\author{M.~Orlita}
\affiliation{LNCMI-CNRS (UJF, UPS, INSA), BP 166, 38042 Grenoble
Cedex 9, France} \affiliation{Institute of Physics, Charles
University, Ke Karlovu 5, 121 16 Praha 2, Czech Republic}
\author{Yu.~I.~Latyshev}
\affiliation{Institute of Radio Engineering and Electronics, RAS,
Mokhovaya 11-7, 125009, Moscow, Russia}
\author{M.~Potemski}
\affiliation{LNCMI-CNRS (UJF, UPS, INSA), BP 166, 38042 Grenoble
Cedex 9, France}

\date{\today}

\begin{abstract}
The micro-Raman scattering response of a graphene-like location on
the surface of bulk natural graphite is investigated both at
$T=\unit{4.2}{K}$ and at room temperature in magnetic fields up to
$29$~T. Two different polarization configurations, co-circular and
crossed-circular, are employed in order to determine the Raman
scattering selection rules. Several distinct series of electronic
excitations are observed and we discuss their characteristic
shapes and amplitudes. In particular, we report a clear splitting
of the signals associated with the inter-Landau level excitations
$-n\rightarrow+n$. Furthermore, we observe the pronounced
interaction of the zone-center E$_{\text{2g}}$-phonon with three
different sets of electronic excitations. Possible origins for
these graphene-like inclusions on the surface of bulk graphite are
discussed.
\end{abstract}

 \maketitle

\section{Introduction}

Raman scattering is a powerful technique that provides information
concerning single particle and collective excitations in
solids.~\cite{LSIS1} Phonon excitations in graphene based systems
have been widely studied in the last years~\cite{Jorio2011}
providing a comprehensive understanding of the graphene phonon
band structure,~\cite{Ferrari2006,Mafra2007} of the different
Raman scattering mechanisms and of the effect of the stacking on
the observed spectrum in the case of multilayer graphene
specimens.~\cite{Faugeras2008} This technique has recently been
extended to address electronic excitations in carbon based
systems.

Indirect signs of electronic excitations have first been observed
on the evolution of the the $E_{2g}$ phonon feature (G band
feature) as a function of some specific thermodynamical parameter.
The energy of the G band feature is mainly determined by the
strong C-C bond and by the graphene crystal hexagonal symmetry,
but also by the electron-phonon interaction, which can be tuned by
changing the Fermi
energy~\cite{Ando2006,Lazzeri2006b,Pisana2007,Yan2007a,Das2008} or
by modifying the electronic density of states by applying a
magnetic field perpendicular to the plane of a graphene crystal.
The $B=0$ monotonic electronic excitation spectrum then transforms
into excitations among discrete Landau levels (LL), with energies
$E_{\pm n}=\pm v_F\sqrt{2e\hbar Bn}$ with $n=0,1,2\dots$. This LL
spectrum is determined by a single parameter, the Fermi velocity
$v_F$. By increasing the magnetic field intensity, inter Landau
level excitations can be tuned in resonance with the phonon which
displays the so-called magneto-phonon
effect.~\cite{Ando2007a,Goerbig2007,Faugeras2009,Yan2010,Faugeras2011}

Raman scattering from electronic excitations have been directly
observed in metallic single wall carbon
nanotubes~\cite{Farhat2011}, in bulk
graphite~\cite{GarciaFlores2009,Kossacki2011,KimPRB12} and in
graphene-like inclusions on the surface of bulk
graphite.\cite{Faugeras2011} Raman scattering techniques combined
with high magnetic fields are now established as being an
extremely well adapted tool to perform the Landau level
spectroscopy, to explore the details of the electronic excitation
spectrum and of the electron-phonon coupling in these systems.
Such experiments are performed in the visible range of energy and
benefit from the particularly efficient focusing and polarization
optics that are well developed in this range of energy.

In this paper, we report on a polarization resolved magneto-Raman
scattering experiment performed on specific locations on the
surface of bulk graphite which show a number of characteristic
attributes of graphene. This work is an extension of our previous
experiments~\cite{Faugeras2011}, addressing the details of
magneto-Raman scattering properties of such graphene-like
locations. We interpret our results in line with other
experiments~\cite{Li2009,Neugebauer2009} pointing towards graphene
layers decoupled from the sequence of Bernal-stacked graphene
layers forming bulk graphite. Low energy magneto-optical
absorption experiments performed on similar domains indicate that
they are quasi neutral with a carrier density of $\sim
10^9$~cm$^{-2}$ and they are characterized by an ultra high
electronic mobility exceeding
$\unit{10^7}{cm^2V^{-1}s^{-1}}$.\cite{Neugebauer2009} We focus on
the electronic Raman scattering signals and their evolution with
increasing magnetic fields. Our low temperature, polarization
resolved measurements bring direct information concerning (i) the
symmetry properties of the observed excitations, (ii) the
electron-hole asymmetry and (iii) the different manifestations of
the electron-phonon interaction in this graphene-based system.
Interestingly, these properties can be traced even at room
temperature. Despite the fact that graphene-like inclusions on the
surface of bulk graphite and graphene share many properties, our
detailed study of electronic excitations and their evolution with
increasing magnetic field reveals a number of striking differences
indicating a non-negligible interaction with the underlying
graphite substrate inducing Landau level mixing.

The paper is organized as follows: we start by introducing the
relevant inelastic light scattering selection rules in sec.
\ref{sec:1}. We continue with a description of our experiment in
sec. \ref{sec:2}, followed by a discussion of our experimental
results in secs. \ref{sec:3} and \ref{sec:4}. A summary and
conclusion is given in sec. \ref{sec:6}.

\section{Theory} \label{sec:1}

One of the main advantages of Raman scattering spectroscopy for
the study of low energy excitations in solids, is that it is
performed in the visible range of energy allowing for the use of
focusing optics. Physical systems can hence be probed locally (on
the $\mu m$ scale), in contrast to conventional methods of
infrared magneto-spectroscopy. It also allows us to use broad-band
polarizing optical elements which are largely developed in the
visible range of energy. Such polarized Raman scattering
experiment bring some fundamental information about the symmetry
of the probed excitations. Here, we consider the quasi
back-scattering Faraday geometry (back scattering geometry with
the use of a high numerical aperture lens providing many different
angles for the incident and collected photons) and the magnetic
field is applied perpendicular to the plane of the 2D crystal. In
the case of graphene subjected to a high magnetic field, the main
observable excitations are optical phonons and inter Landau level
electronic excitations.

Raman scattering selection rules concerning optical phonons are
today well established. The Raman G band is a first-order Raman
scattering process involving a doubly degenerate E$_{\text{2g}}$
optical phonon at the $\Gamma$ point of the phonon Brillouin zone.
These phonons carry an angular momentum of $\pm 1$, are observed
in the crossed circular polarization configurations
$\sigma\pm/\sigma\mp$ (incident polarization/outgoing
polarization) corresponding to an angular momentum transfer of
$\pm2$, and are completely suppressed in the co-circular
polarization configurations
$\sigma\pm/\sigma\pm$.~\cite{Faugeras2011,Kossacki2011} As
discussed in Ref.~\onlinecite{Basko2008,Basko2009,Kossacki2011},
this phonon is seen through a \textit{weakly allowed} Raman
scattering process due to the trigonal warping that provides an
additional angular momentum transfer of $\pm3$.

\begin{figure}
\includegraphics[width=0.47\textwidth]{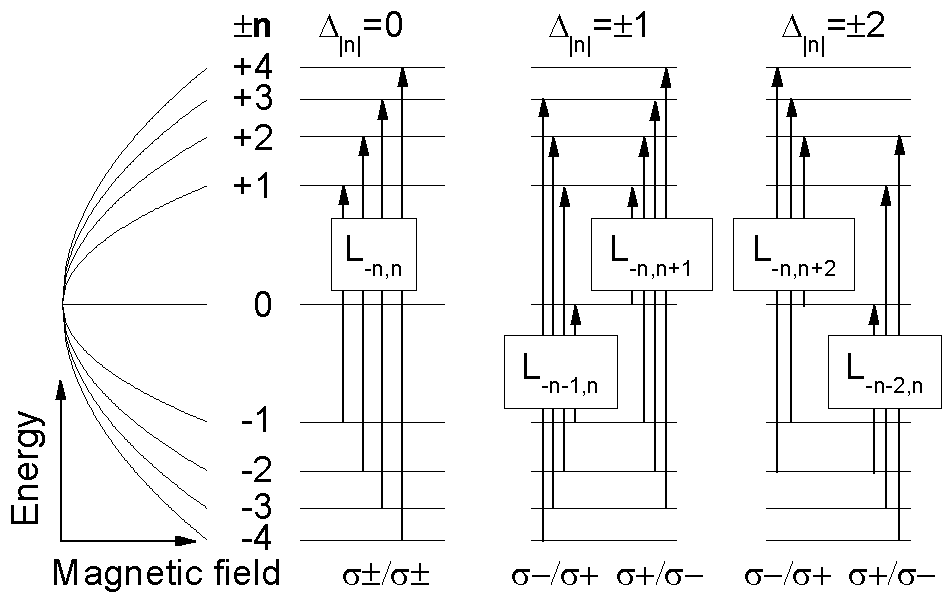}
\caption{\label{fig:selrules} Schematic graphene LL spectrum with
three types of electronic excitations. Selection rules for the
inelastic light scattering processes are indicated.}
\end{figure}

Electronic Raman scattering in graphene subjected to a quantizing
magnetic field involves inter-LL excitations. Raman scattering
selection rules for electronic excitations have been derived
theoretically for graphene~\cite{Kashuba2009} and for bilayer
graphene~\cite{Mucha-Kruczynski2010}. In Fig. \ref{fig:selrules},
we present the graphene LL fan chart and we distinguish three
types of excitations according to the change in the absolute value
of the LL index $\Delta |n|=|n_f|-|n_i|$ where $n_f$ ($n_i$) is
the index of the final (initial) LL. We label excitations
$n_i\rightarrow n_f$ as $L_{n_i,n_f}$ and $\hbar\Delta |n|$ is the
angular momentum transferred from/to the electronic system during
the scattering process. $\Delta |n|=\pm2$ excitations are expected
to be active in the crossed-circular polarization configuration
$\sigma\mp/\sigma\pm$, and $\Delta |n|=\pm0$ excitations in the
co-circular polarization configuration
$\sigma\pm/\sigma\pm$.~\cite{Kashuba2009}

Similar to the case of the $E_{2g}$ phonon, scattering from
$\Delta |n|=\pm1$ excitations (optical-like excitations) is
forbidden in the isotropic approximation but becomes allowed due
to the trigonal warping that provides an additional angular
momentum transfer of $\pm3$.~\cite{Kashuba2009} These excitations
are hence expected to be observed in the $\sigma\pm/\sigma\mp$
polarization configuration. Kashuba and Fal'ko~\cite{Kashuba2009}
find that the most pronounced signals from electronic Raman
scattering in graphene should be due to the $\Delta |n|=0$
excitations. Furthermore, the associated signal intensities are
expected to scale $\propto B$.~\cite{Kashuba2009}

Interactions between electronic excitations and the
E$_{\text{2g}}$ phonon leads to a magnetic-field dependent
renormalization of the phonon energy and to pronounced avoided
crossings of the coupled phonon- and electronic
modes.\cite{Ando2007a,Goerbig2007,Faugeras2009,Yan2010,Faugeras2011}
Relevant electronic excitations for this effect are the $\Delta
|n|=\pm1$ excitations.\cite{Kashuba2009} However, recent
experimental results~\cite{Faugeras2011} suggest a violation of
this selection rule in graphene-like locations found on the
surface of natural graphite as $\Delta |n|=0$ excitations were
also shown to strongly interact with the phonon.

\section{Experiment and Results} \label{sec:2}

We have performed magneto-Raman scattering experiments in the
quasi back-scattering Faraday configuration. The excitation laser
beam, provided by a Ti:sapphire laser setup, was tuned to
$\approx\lambda=\unit{785}{nm}$.  Two optical fibers with core
diameters of $\unit{5}{\micro m}$ and $\unit{200}{\micro m}$ were
used respectively to excite the sample and to collect the
scattered light. An aspherical lens is used to focus the
excitation laser to a spot of $\sim\unit{1}{\micro m}$ diameter.
The collected signal was dispersed in a single grating
spectrometer equipped with a liquid nitrogen-cooled CCD detector.
The optical power on the sample was of the order of
$\unit{7}{mW}$. The sample was mounted on x-y-z piezo-stages,
immersed in a low-pressure He atmosphere kept at $T=\unit{4.2}{K}$
(except for room temperature experiments) and placed in the center
of a resistive magnet delivering continuous magnetic fields up to
$\unit{29}{T}$. A set of optical filters was used to minimize the
contribution of superfluous light (Raman signals from the fibers,
stray light etc.) to the spectra. Experiments have been performed
both in the co-circular polarization configurations
($\sigma\pm/\sigma\pm$) and in the crossed-circular polarization
configurations ($\sigma\pm/\sigma\mp$). Both pairs of circular
polarization configuration, e.g., $\sigma+/\sigma+$ and
$\sigma-/\sigma-$ in the case of the co-circular configuration,
were achieved by inverting the polarity of the magnetic field with
respect to the light propagation direction.

\begin{figure*}[t]
\includegraphics[width=0.9\textwidth]{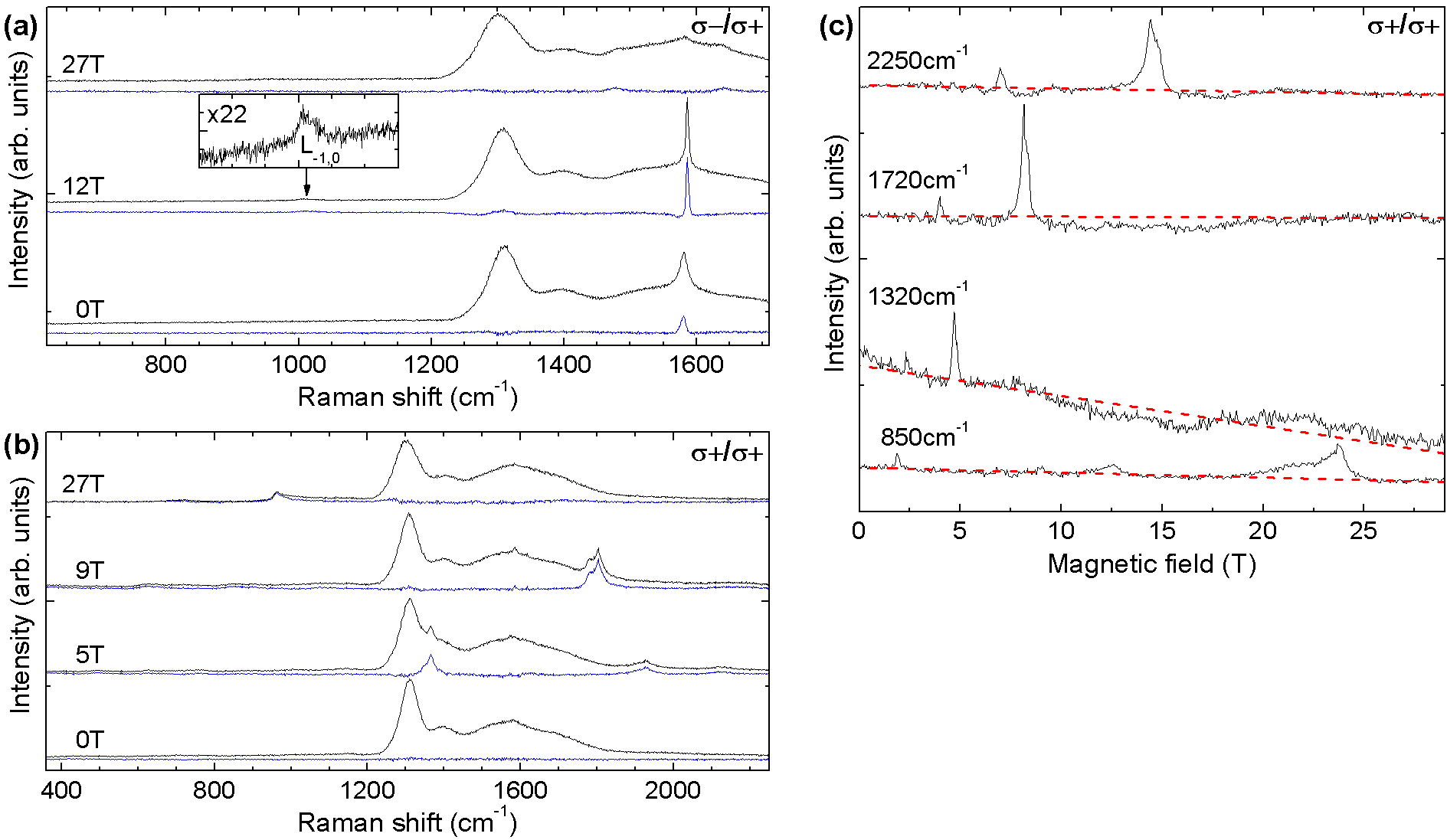}
\caption{\label{fig:rawdata} (Color online) Raw data spectra
(black lines) measured at different values of the magnetic field
in the (a) crossed-circular configuration and (b) co-circular
configuration. Corrected spectra (blue lines in (a) and (b)) are
obtained by subtracting a linear background (red dashed lines in
(c)) from the data at each energy value.}
\end{figure*}

The sample used in this experiment is a piece of natural graphite.
Graphene-like locations are identified on the surface of bulk
graphite using the method described in
Ref.~\onlinecite{Faugeras2011}. Accordingly, we laterally map the
Raman scattering response of the sample subjected to a finite
magnetic field. The field strength is chosen so that the G band
energy is significantly modified with respect to its
$B=\unit{0}{T}$ energy due to the electron-phonon interaction.
This method is straightforward and unambiguous especially since
the amplitude of the energy renormalization of the G band in
graphite is much smaller than in graphene, because of the 3D
Landau bands in the underlying graphite
substrate.\cite{Kossacki2011}

In Fig.~\ref{fig:rawdata} we show several raw spectra (black
lines) of the magneto-Raman scattering response of such a
graphene-like location as measured in the two polarization
configurations at $T=\unit{4.2}{K}$. Apart from a relatively flat
background, these spectra are affected by a parasitic signal
extending from $\sim\unit{1200}{cm^{-1}}$ to
$\sim\unit{1900}{cm^{-1}}$ possibly due to inelastic scattering or
luminescence contributions from one of the optical elements or
optical fibers used in the set-up. However, since this signal only
shows a weak, and to a first approximation linear, dependence on
the magnetic field, we can correct our spectra by subtracting, for
each energy, a linear in magnetic field background (red dashed
lines in Fig. \ref{fig:rawdata}(c)). The result of such a
correction is shown as blue lines in Fig. \ref{fig:rawdata}). in
the following, we mainly take advantage of the better contrast
obtained through this correction in order to plot intensity false
color maps, while we use uncorrected spectra to discuss the
details of line shapes and intensities.

\begin{figure*}[]
\includegraphics[width=\textwidth]{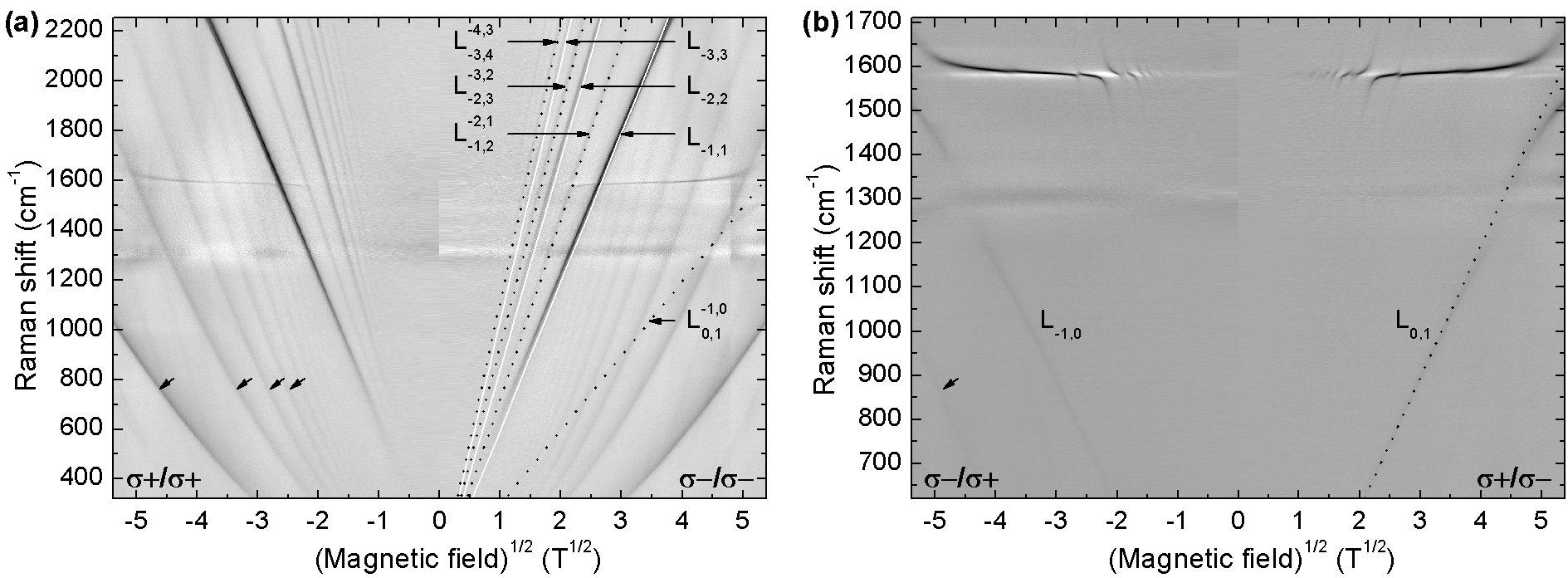}
\caption{\label{fig:overview} Magneto-Raman scattering response of
graphene on natural graphite measured in (a) co-circular
configuration and (b) crossed-circular configuration: Intensity
false color plots of the background-corrected spectra. Black
(white) corresponds to high (low) intensity.}
\includegraphics[width=0.9\textwidth]{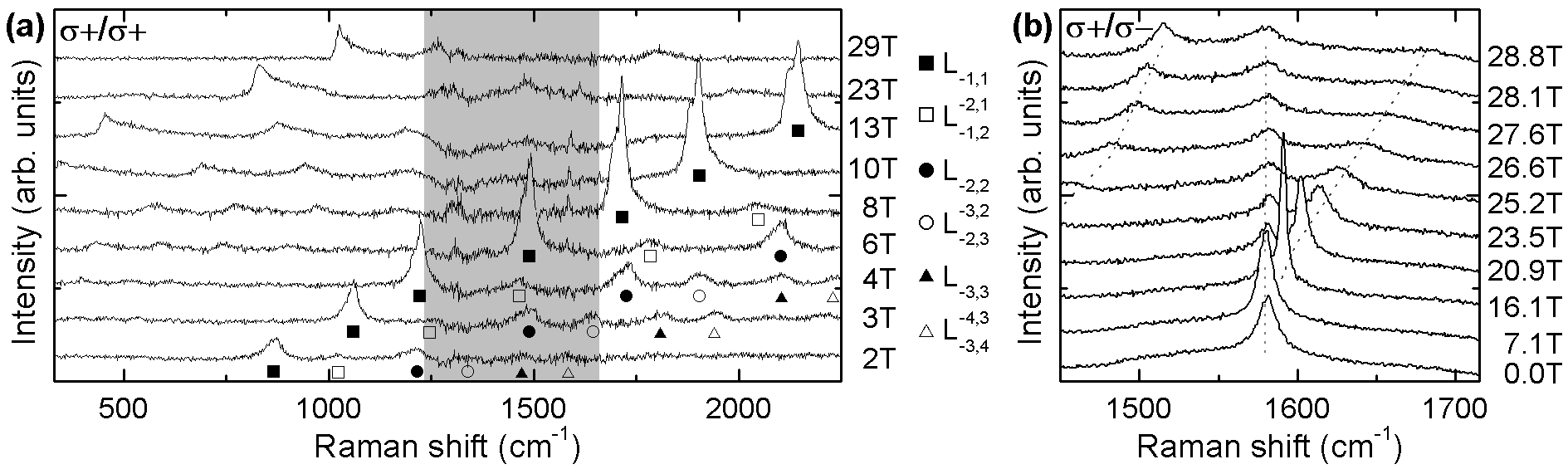}
\caption{\label{fig:Mspec} Magneto-Raman scattering response of
graphene on natural graphite measured in (a) co-circular
configuration and (b) crossed-circular configuration. The spectra
in (a) are background-corrected; symbols identify signals that are
due to electronic excitations in graphene. Intensities in the gray
shaded region are slightly affected by the background correction
described in the text. Dotted lines indicate three Raman G band
components in (b).}
\end{figure*}

An overview of our experimental results is shown in Fig.
\ref{fig:overview}. Figs. \ref{fig:overview}(a) and
\ref{fig:overview}(b) are intensity false color maps of the
background corrected spectra measured in the co-circular
configuration and in the crossed-circular configuration,
respectively, as a function of $\sqrt{B}$. Correspondingly, we
present some individual spectra without any background subtraction
in Fig. \ref{fig:Mspec}(a) and with background subtraction in Fig.
\ref{fig:Mspec}(b) slightly affecting the intensities in the gray
shaded region. Figs. \ref{fig:overview} and \ref{fig:Mspec}
clearly demonstrate that various magnetic field-dependent signals
characterized by different line shapes are observed in the Raman
scattering response of our sample, and that these different
signals are selected by the two distinct polarization
configurations. Similar results have been obtained on $4$
different pieces of natural graphite and on HOPG. We limit all
discussions in this paper to the graphene-like excitations. The
Raman scattering response from the underneath graphite substrate,
which is also observed when placing the laser spot on a
graphene-like domain and which is indicated by the diagonal arrows
in Fig. \ref{fig:overview}, has been discussed in details in
Ref.~\onlinecite{Kossacki2011}.

\section{Co-circular configuration} \label{sec:3}

We present in Fig. \ref{fig:overview}(a) a false color map of the
scattered intensity as a function of the magnetic field measured
in the co-circular polarization configurations. Several series of
magnetic field dependent features can be identified. Their
characteristic, in a first approximation, $\sqrt{B}$ evolution
indicates that they arise from electronic excitation among
graphene LLs. The most pronounced features correspond to inter-LL
excitations of the type L$_{-n,n}$. Their evolution with
increasing magnetic fields can be described (full white lines in
Fig. \ref{fig:overview}(a)) by the graphene LL spectrum given by
$E_{\pm n}=\pm v_F\sqrt{2e\hbar Bn}$ with $n=0,1,2\dots$ and a
Fermi velocity $v_F=\unit{(1.04 \pm 0.03)\cdot 10^6}{m\cdot
s^{-1}}$. Results obtained in the $\sigma+/\sigma+$ and in the
$\sigma-/\sigma-$ configurations are identical within the
resolution of our experiment. Our measurement further confirms
that inelastic light scattering involving this kind of electronic
excitations is allowed in the co-circular configuration, as
expected from the zero angular momentum transfer.

\begin{figure*}[]
\includegraphics[width=0.9\textwidth]{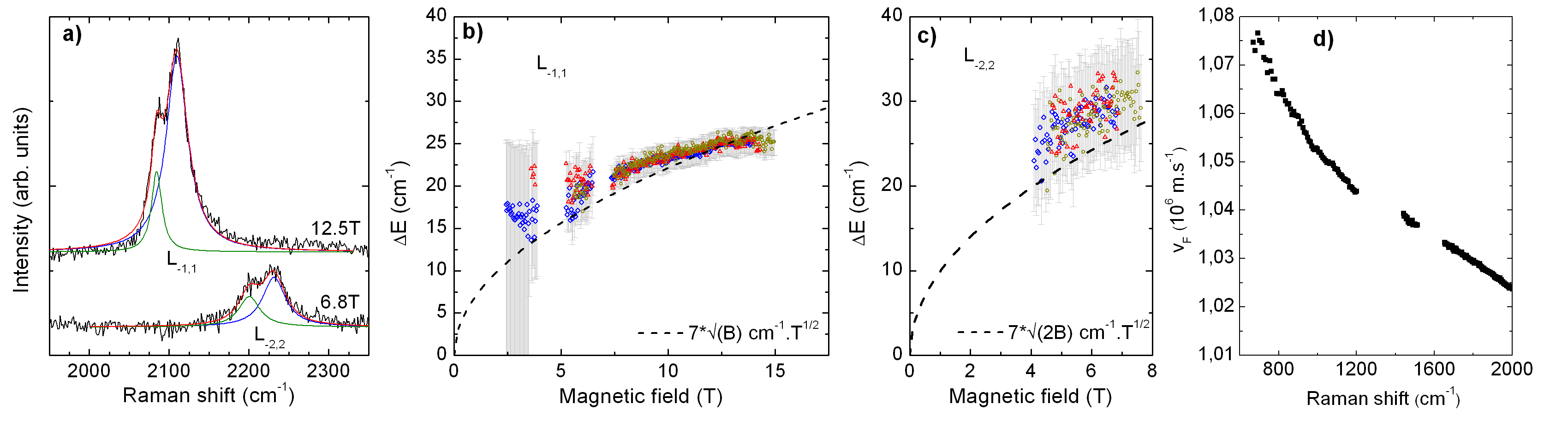}
\caption{\label{fig:COCsignals} (Color online) (a) Measured Raman
scattering response at two values of $B$. The signals associated
with the excitations L$_{-1,1}$ and L$_{-2,2}$ are fitted with two
Lorentzian functions. The split between the centers of these two
Lorentzians $\Delta E$ is shown in (b) and (c) as extracted from
three data sets acquired by measuring the same sample. d) Fermi
velocity for the main component of the L$_{-1,1}$ as a function of
the energy of the excitation.}
\end{figure*}

However, a detailed analysis of these Raman scattering spectra
reveals that to properly describe the magnetic field evolution of
the $L_{-n,n}$ series, an energy dependent $v_F$ has to be used.
This is particularly visible in the case of the $L_{-1,1}$ feature
in Fig.~\ref{fig:overview}(a). As shown in
Fig.~\ref{fig:COCsignals}(d), the Fermi velocity $v_F$ decreases
by $\sim 5\%$ as the energy of $L_{-1,1}$ shifts from
$\unit{110}{meV}$ to $\unit{280}{meV}$. Recent magneto-transport
experiments performed with suspended, high mobility graphene
specimens have revealed an electron-electron interactions induced
increase of the Fermi velocity, of nearly a factor $3$, as a
function of the Fermi energy~\cite{Elias2011}, in agreement with
theoretical predictions.~\cite{Gonzales1994,Gonzales1999} The
graphene system studied here shows an increase of $\sim 5\%$ of
the Fermi velocity that could be strongly reduced, in this frame,
by the underlying graphite substrate efficiently screening Coulomb
interaction. In contrast to Ref.~\onlinecite{Elias2011}, our
magneto-Raman scattering experiment probes the evolution of the
different electronic states versus energy at a fixed Fermi energy.

\begin{figure}[]
\includegraphics[width=0.4\textwidth]{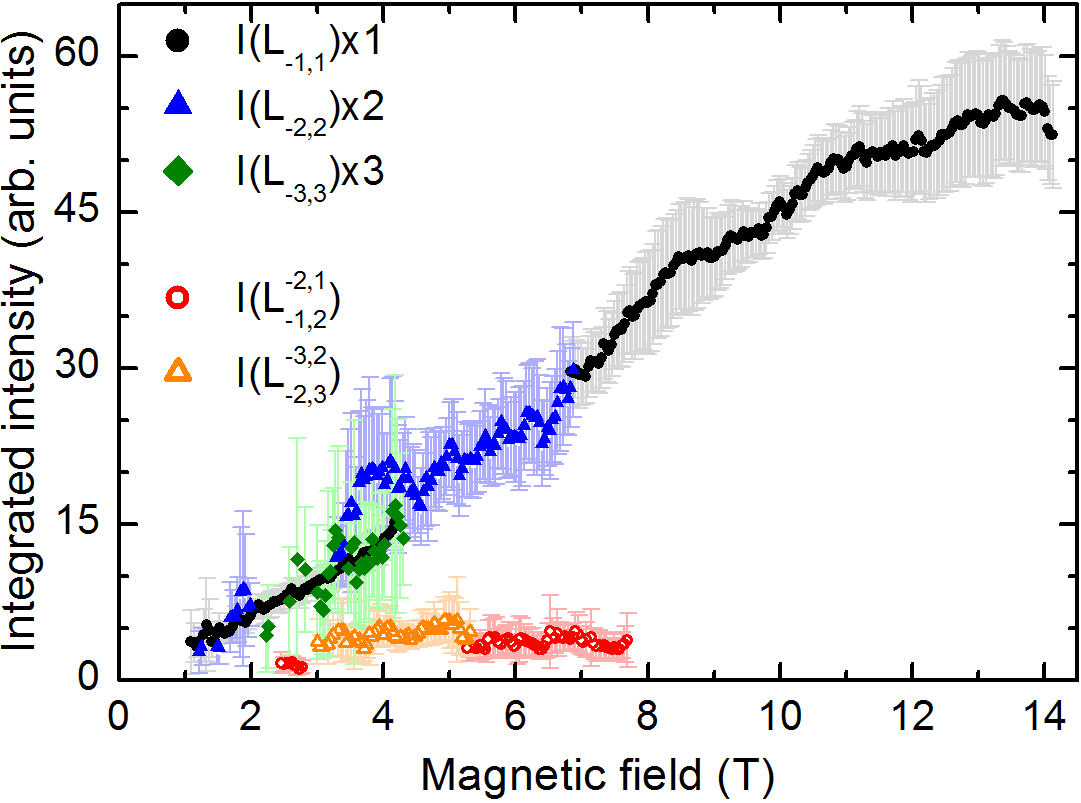}
\caption{\label{fig:COCintint} (Color online) Integrated
intensities of the signals identified in Fig.
\ref{fig:overview}(b) extracted from the data acquired in
$\sigma-/\sigma-$ configuration.}
\end{figure}

As can be seen in Fig. \ref{fig:Mspec}(a), the amplitudes of the
different L$_{-n,n}$ excitations clearly depend on both the
magnetic field and on the LL index $n$. At a given value of the
magnetic field, the frequency-integrated intensity
$I(\text{L}_{-1,1})$ is bigger than $I(\text{L}_{-2,2})$ which is
in turn bigger than $I(\text{L}_{-3,3})$. The integrated
intensities of these three excitations, extracted from the
measured spectra and corrected for the wavelength dependence of
the quantum efficiency of our CCD camera, are shown in Fig.
\ref{fig:COCintint}. As can be seen in this Figure, the intensity
of each $L_{-n,n}$ transition scales $\propto B$ in agreement with
theory~\cite{Kashuba2009}. In addition, we find that
$I(\text{L}_{-n,n})$ scales $\propto 1/n$. This latter fact is
emphasized in Fig.~\ref{fig:COCintint} by multiplying all
$I(\text{L}_{-n,n})$ by $n$.

Furthermore, the observed L$_{-n,n}$ excitations show strongly
asymmetric line shapes, see Fig. \ref{fig:Mspec}(a). As shown in
Fig. \ref{fig:COCsignals}(a), it appears clearly that, at high
magnetic fields, both $L_{-1,1}$ and $L_{-2,2}$ signals are
composed of two components with one of lower amplitude at the
low-energy side. This characteristic is the same in both
$\sigma+/\sigma+$ and $\sigma-/\sigma-$ configuration. As
demonstrated in Fig. \ref{fig:COCsignals}(a), we quantify this
phenomenon by fitting each of these L$_{-n,n}$ excitations with
the sum of two Lorentzian functions. The splitting energy $\Delta
E$ between the centers of the two individual Lorentzians, can then
be extracted as is shown in Figs. \ref{fig:COCsignals}(b) and
\ref{fig:COCsignals}(c) for L$_{-1,1}$ and L$_{-2,2}$,
respectively. The splits are extracted from three different sets
of data acquired during three consecutive measurements (two
performed in $\sigma+/\sigma+$ configuration and one in
$\sigma-/\sigma-$ configuration) on the same spot on the sample,
to ensure reproducibility of the effect.

As can be seen in Fig.~\ref{fig:COCsignals}, the magnitude of the
extracted splits increases with increasing magnetic fields.
Moreover, at a given value of $B$, $\Delta E$ is bigger for
L$_{-2,2}$ than for L$_{-1,1}$. $\Delta E$ shows a
$\sqrt{B}$-dependence, indicated by the black dashed lines in
Figs.~\ref{fig:COCsignals}(b) and (c). Such a magnetic
field-dependence could in general be consistent with the existence
of two cones with different Fermi velocities $v_F1$ and $v_F2$. In
such a case, the splitting of L$_{-n,n}$ would be $\Delta E=\Delta
v_F\cdot2\sqrt{2e\hbar Bn}$, where $\Delta v_F=|v_F1-v_F2|\approx
\unit{0.06\cdot10^6}{m/s}$. The magnitude of $\Delta E$ would
consequently depend on the LL index $n$, which we have taken into
account for drawing the black dashed lines in Figs.
\ref{fig:COCsignals}(b) and (c). Although the physical reason for
the appearance of the observed $\Delta E$ is not fully understood,
we believe that it may be induced by a residual interaction with
the underlying graphite substrate. The recent report of an equally
unusual LL splitting induced by the twist in a graphene bilayer
maybe points in the same direction.~\cite{deGail2011}

We now turn to the observation of a set of $\sqrt{B}$-dependent
features that we assigned in Fig. \ref{fig:overview}(a) to
L$_{-n,n+1}^{-(n+1),n}$ excitations (a regrouped form accounting
for L$_{-n,n+1}$ and L$_{-(n+1),n}$). As was discussed in
Sec.~\ref{sec:1}, inelastic light scattering involving this type
of excitations is not theoretically expected in this polarization
configuration.\cite{Kashuba2009} Nevertheless, we find a good
agreement with our data when calculating their energies (dotted
black lines) again using $v_F=\unit{(1.04 \pm
0.03)\cdot10^6}{m/s}$. These features are pronounced even far away
in energy from the Raman G band only for $n\ge1$ which appear in
our spectra at low values of the magnetic field,
$B\le\unit{10}{T}$. The L$_{0,1}^{-1,0}$ excitation is much weaker
and the associated signal can hardly be identified in the spectra
(a fact especially clear for $B\ge\unit{10}{T}$).
Fig.~\ref{fig:COCintint} shows the integrated intensities of the
L$_{-1,2}^{-2,1}$ and L$_{-2,3}^{-3,2}$ excitations. In comparison
to the L$_{-n,n}$ series, the observed behavior for the
L$_{-n,n+1}^{-(n+1),n}$ series is much less pronounced. The
dependence of their oscillator strengths on $B$ is certainly weak
and a decrease with decreasing $n$ could be suggested.

\begin{figure}
\includegraphics[width=0.8\columnwidth]{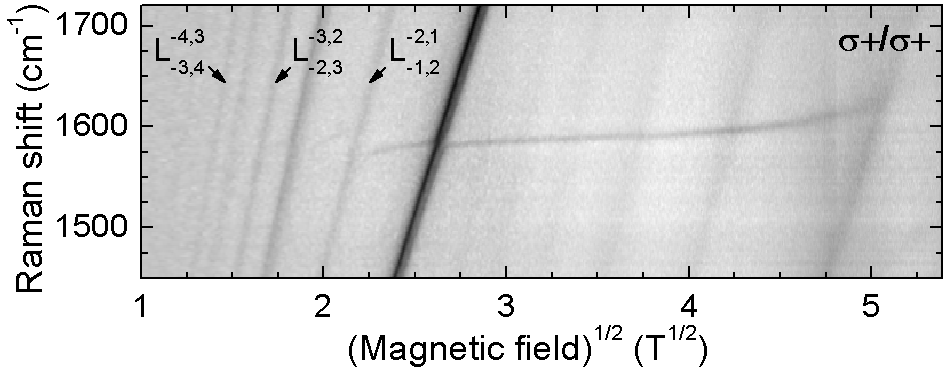}
\caption{\label{fig:COCepsart} Raman scattering response as a
function of the magnetic field: intensity false color plot as in
Fig. \ref{fig:overview}(a).}
\end{figure}

A remarkable aspect of the L$_{-n,n+1}^{-n-1,n}$ excitations
observed in the co-circular polarization configuration is that
they do not interact with the E$_{\text{2g}}$-phonon. This is
clearly seen in Fig.~\ref{fig:COCepsart} as the magnetic field
dependence of the Raman signals associated with
L$_{-n,n+1}^{-n-1,n}$ form straight lines instead of the expected
avoided crossings with the phonon (compare with the weak signal
from the Raman G band, visible due to a polarization leak). We
therefore believe that these particular electronic excitations
observed here do not have the same symmetry as the phonon, but
correspond rather to the complementary set of usually
infrared-active modes that are characterized by the irreducible
representation (IR) E$_{\text{1u}}$ of D$_{\text{6h}}$. The origin
of asymmetric transitions (optical-like) which are Raman active
and seen in the co-circular polarization configuration is not
clear. We may only speculate that their appearance is due to
Landau level mixing (transitions involving higher LL indexes are
more pronounced) and/or their observation is allowed for the
particular graphene system studied here due to some residual
interaction of the graphene flakes with the graphite substrate.
Optionally, we note that those transitions are expected to be
forbidden in Raman scattering processes in case of an ideal, back
scattering Faraday geometry but they are perhaps active in the
present measurement~\footnote{We thank V.I. Fal'ko for pointing
out this possibility} due to an imperfection of our experimental
configuration (large apertures of the excitation and of the
collected light).

\begin{figure}[]
\includegraphics[width=0.47\textwidth]{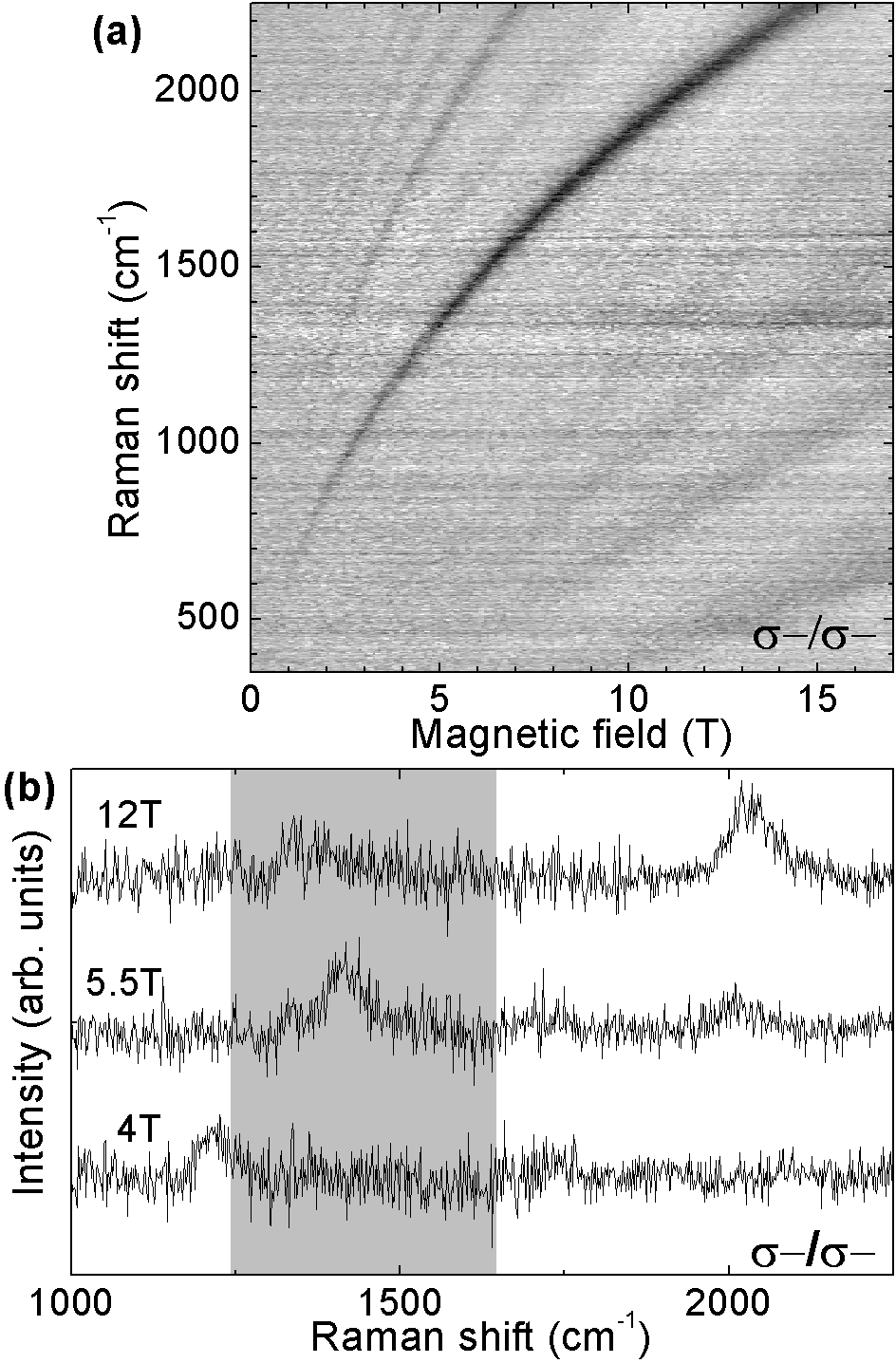}
\caption{\label{fig:COCroomtemp} Room temperature magneto-Raman
scattering response in co-circular configuration: (a) intensity
false color plot where black (white) corresponds to high (low)
intensity and (b) individual, background-corrected spectra.}
\end{figure}

Finally, we measured the room temperature magneto-Raman scattering
response of a similar graphene location on the surface of bulk
natural graphite. In Fig. \ref{fig:COCroomtemp} we show the
obtained spectra after subtraction of the zero-field scattering
response in the form of an intensity false color plot.
Surprisingly, although the magnetic field resolution is worse than
in our low-temperature experiment, one can clearly see that the
same physics is observed. Notably, Raman scattering from
electronic excitations in graphene, discussed above, is still
clearly measured up to room temperature. The intensities of the
different excitations also show a behavior comparable to the one
observed at $T=\unit{4.2}{K}$.

\section{Crossed-circular configuration} \label{sec:4}

\begin{figure*}[t]
\includegraphics[width=0.95\textwidth]{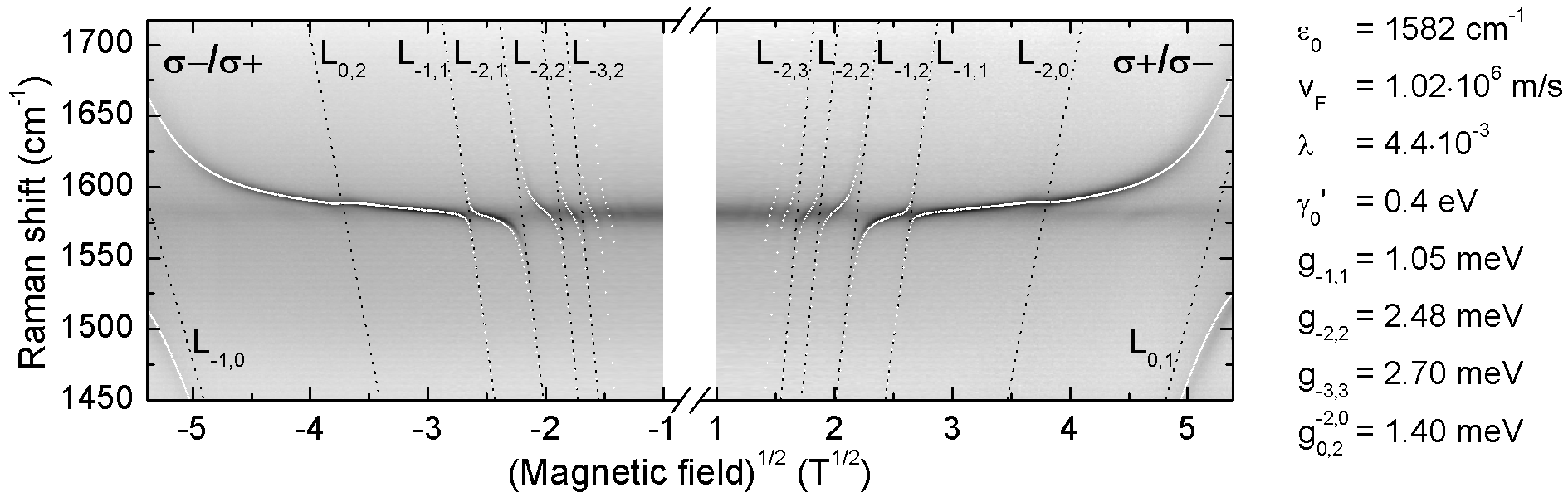}
\caption{\label{fig:CRCE2g} Magneto-Raman scattering response:
Intensity false color plot where black (white) corresponds to high
(low) intensity. Black dotted lines are the energies of the
inter-LL excitations that interact with the zone-center
E$_{\text{2g}}$-phonon. White dots are solutions to
Eq.~\eqref{eq:EPC} calculated with the parameters given on the
right.}
\end{figure*}

Figs. \ref{fig:overview}(b) and \ref{fig:Mspec}(b) show the Raman
scattering response of the graphene-like location measured in the
crossed-circular configuration. In contrast to the co-circular
configuration, the G band feature is observed~\cite{Kossacki2011}.
When applying a magnetic field, the magneto-phonon
effect~\cite{Ando2007a,Goerbig2007} appears as an avoided crossing
behavior each time the energy of specific inter-LL excitations is
tuned in resonance with the phonon. As discussed further below
(see Fig. \ref{fig:CRCE2g}), the most pronounced of these avoided
crossings are caused by electronic excitations of the
L$_{-n,n+1}^{-(n+1),n}$ series. The raw spectra in Fig.
\ref{fig:Mspec}(b) illustrate well the magnitude of this effect
for the L$_{0,1}$ excitation. These spectra, however, are
resolution limited in energy. The acquisition of some
high-resolution spectra showed that the full width at half maximum
of the G band (FWHM$_{\text{G}}$) was in general much smaller than
the one suggested in Fig. \ref{fig:Mspec}(b), as low as e.g.
$\text{FWHM}_{\text{G}}\approx\unit{2.5}{cm^{-1}}$ measured at
$B=\unit{11}{T}$.

It should be noted, that L$_{0,1}^{-1,0}$ is in fact the only
electronic excitation from which we observe inelastic light
scattering, in this polarization configuration, also far away in
energy from the Raman G band (see Fig. \ref{fig:overview}(b) and
the inset in Fig. \ref{fig:rawdata}(a)). Other electronic
excitations of the L$_{-n,n+2}^{-(n+2),n}$ series, expected to be
active in this polarization configuration,\cite{Kashuba2009} are
probably too weak to be detected in our experimental setup. As
shown in Fig. \ref{fig:overview}(b), apart from the
L$_{0,1}^{-1,0}$ excitations, we only observe traces of Raman
scattering from other electronic excitations of the
L$_{-n,n+1}^{-(n+1),n}$ series close to the phonon energy with
which they hybridize.

\begin{figure}
\includegraphics[width=0.47\textwidth]{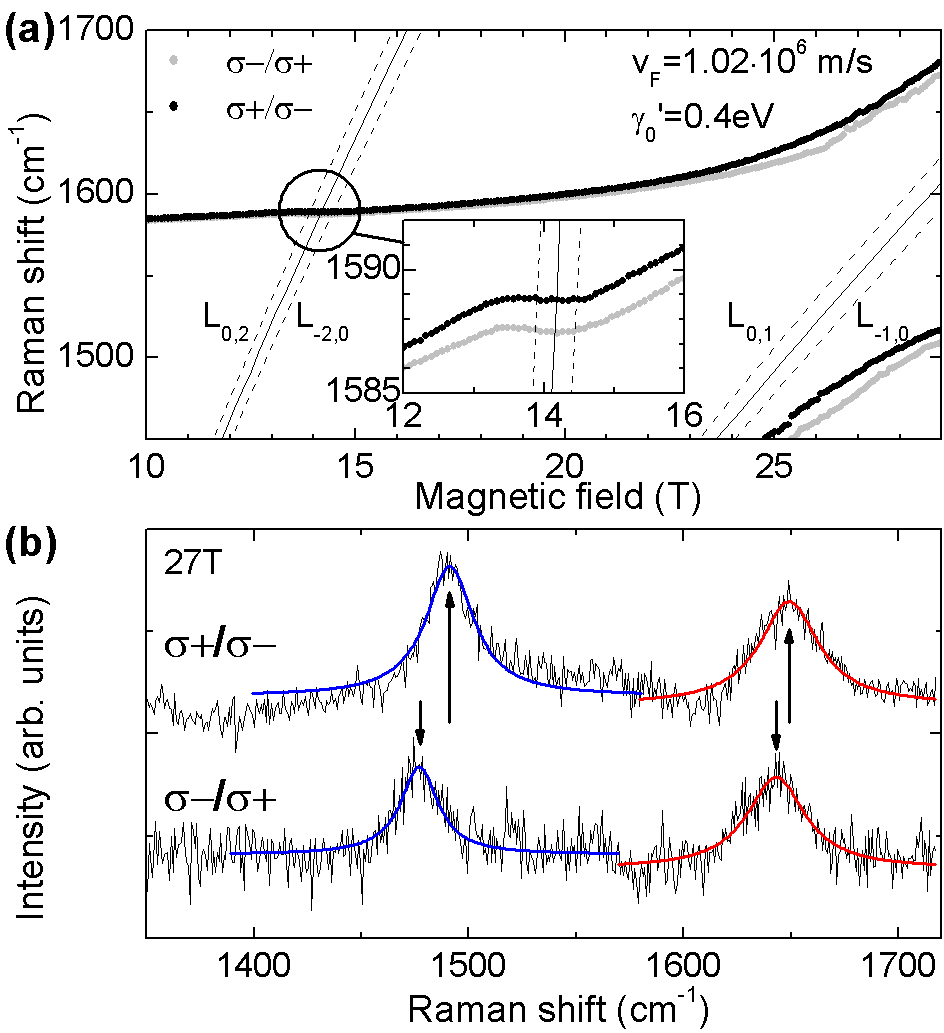}
\caption{\label{fig:asym} (a) Energy positions of the Raman G band
in $\sigma+/\sigma-$ and $\sigma-/\sigma+$ extracted from the
background-corrected spectra by doing single-Lorentzian fits as
demonstrated in (b). Energies of some electronic excitations
(dashed lines) corrected by a finite $\gamma_0'$ as compared to
the case of $\gamma_0'=0$ (full lines, degenerate excitations).}
\end{figure}

Strikingly, the magnetic field dependence of the Raman G band is
not symmetric with respect to inversion of the polarization
configuration. The differences in the energy of the signal between
$\sigma+/\sigma-$ and $\sigma-/\sigma+$ appear especially at
higher magnetic field strengths, see Fig. \ref{fig:overview}(b).
To emphasize this fact, we fitted the Raman G band components with
a single Lorentzian in order to extract their energy positions as
shown in Fig. \ref{fig:asym}(b). The result is plotted for both
$\sigma+/\sigma-$ and $\sigma-/\sigma+$ in Fig. \ref{fig:asym}(a),
clearly showing the mentioned differences which reflect, as we
argue, an asymmetry between electron and hole states. Finite
doping could also induce differences between $\sigma+/\sigma-$ and
$\sigma-/\sigma+$, however, not changing the value of the Fermi
velocity by rather the magnitude of the observed
anticrossings~\cite{Goerbig2007}. This asymmetry is caused by the
trigonal warping, which in graphene is described by a second-order
term in the expansion of the graphene LL energies, determined by
the next-nearest-neighbor hopping integral $\gamma_{0}'$. In this
respect, graphene differs from graphite, where the interlayer
hopping parameter $\gamma_3$ is responsible for the trigonal
warping. Including this term, the energies of the LLs in graphene
are given by $E_{\pm n}=\pm v_F\sqrt{2e\hbar Bn}+9eB/(2\hbar)v_F
\gamma_0' a_0^2n$ where $a_0=\unit{1.42}{\AA}$ is the C-C
distance.\cite{Plochocka2008} Consequently, the energies of
electronic excitations such as L$_{0,1}$ and L$_{-1,0}$ are no
longer degenerated in the presence of this electron-hole
asymmetry. this can be seen in Fig.~\ref{fig:asym}(a) where we
compare the magnetic field dependence of the energy of the
degenerate L$_{0,1}^{-1,0}$ excitation (full black line) with the
corrected energies of L$_{0,1}$ and L$_{-1,0}$ (dashed black
lines) calculated with the parameters given in the figure. As we
show below, the electron-hole asymmetry in this graphene-like
system can be directly determined from the given data.

It should be noted that the oscillations of the Raman G band are
in fact not as simple as theoretically expected. In Refs.
\onlinecite{Ando2007a} and \onlinecite{Goerbig2007}, only
electronic excitations of the type L$_{-n,n+1}^{-(n+1),n}$ are
considered to hybridize with the phonon, in agreement with what is
observed in multi-layer epitaxial graphene on
SiC($000\overline{1}$).\cite{Faugeras2009} However, in addition to
the previously measured avoided crossings caused by excitations of
the L$_{-n,n+1}^{-n-1,n}$ series~\cite{Yan2010, Faugeras2011} and
of the L$_{-n,n}^{\vphantom{n}}$ series~\cite{Faugeras2011} at
similar locations on bulk graphite, we find clear evidence for a
further interaction due to the L$_{0,2}^{-2,0}$ excitation. This
new effect is illustrated in the inset of Fig.~\ref{fig:asym}(a).
Similar to the mechanism invoked for the observability of
electronic Raman scattering from L$_{-n,n+1}^{-n-1,n}$ excitations
in the co-circular configuration, here the additional and
unexpected interaction is possibly caused by the mixing of LL wave
functions. In the absence of a complete theoretical model, we
describe the coupling between the phonon and L$_{-n,n+1}^{-n-1,n}$
excitations in terms of the dimensionless coupling constant
$\lambda$.\cite{Ando2007a,Faugeras2009} For all the other
electronic excitations of energies $\Delta_{-n,m}$, we use
magnetic field-independent effective coupling parameters
$g_{-n,m}$. Thus, we can calculate the energy $\epsilon$ of the
electron-phonon coupled modes by
solving~\cite{Ando2007a,Goerbig2007}

\begin{equation} \label{eq:EPC}
\begin{split}
\epsilon^2 - \epsilon_0^2 - 4\epsilon_0 \sum_{n=0}^{n_c} \frac{\lambda}{2} E_1^2 \left( \frac{\Delta_{-n,n+1}}{\epsilon^2_{\vphantom{-n,m}}-\Delta_{-n,n+1}^2}+\frac{1}{\Delta_{-n,n+1}^{\vphantom{2}}} \right) & \\
  - 4\epsilon_0\sum_{(-n,m)}g_{-n,m}^2\left(\frac{\Delta_{-n,m}}{\epsilon^2_{\vphantom{-n,m}}-\Delta_{-n,m}^2}+\frac{1}{\Delta_{-n,m}^{\vphantom{2}}}\right) & =0,
\end{split}
\end{equation}

where $\epsilon_0$ is the energy of the phonon at $B=\unit{0}{T}$
and where the second sum is over L$_{-1,1}$, L$_{-2,2}$,
L$_{-3,3}$, L$_{0,2}$ and L$_{-2,0}$. Taking into account the
electron-hole asymmetry, only one of the last two excitation is
considered respectively in either $\sigma+/\sigma-$ configuration
or $\sigma-/\sigma+$ configuration. The same holds for the
L$_{-n,n+1}^{-n-1,n}$ excitations.

\begin{figure}[]
\includegraphics[width=0.47\textwidth]{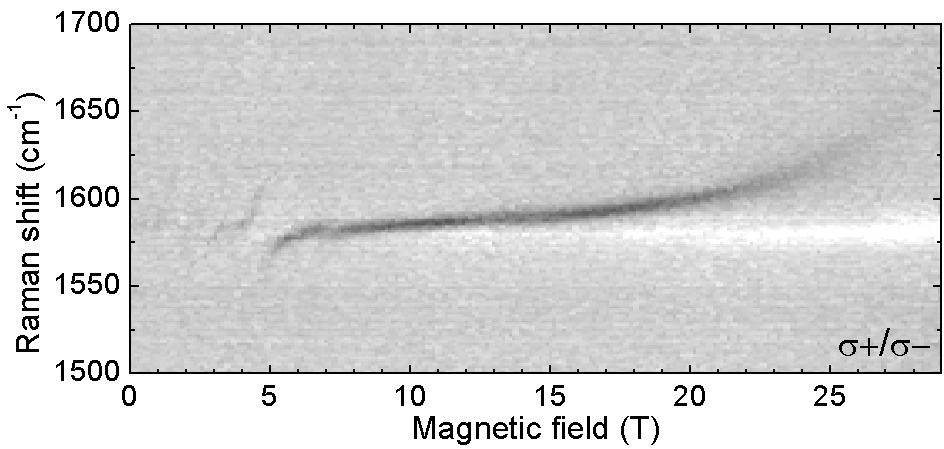}
\caption{\label{fig:CRCroomtemp} Room temperature magneto-Raman
scattering response: Intensity false color plot where black
(white) corresponds to high (low) intensity measured in the
crossed-circular configuration.}
\end{figure}

The best fit to our data, shown in Fig. \ref{fig:CRCE2g}, is
obtained for the parameters given therein.
$\lambda=4.4\cdot10^{-3}$ is found to be in good agreement with
previous experiments on such kind of graphene
samples~\cite{Yan2010, Faugeras2011} as well as on epitaxial
graphene on SiC($000\overline{1}$)~\cite{Faugeras2009}. The
effective coupling strengths $g_{-n,n}$ are found to decrease with
decreasing $n$ as the magneto-phonon resonances take place at
higher values of the magnetic field. This already observed
behavior~\cite{Faugeras2011} is consistent with a mixing of LLs
that decreases with increasing magnetic fields. The parameter
$g_{0,2}^{-2,0}$ is found to be of the same order as the
g$_{-n,n}$. Here, we introduced a non-zero broadening in the
calculation to smear out the avoided crossing caused by
L$_{0,2}^{-2,0}$, since it is not very pronounced. In what
concerns the electron-hole asymmetry, we obtain
$\gamma_0'=\unit{0.4}{eV}$, in agreement with tight-binding
calculations.\cite{Grueneis2008}

Finally, we have performed similar magneto-Raman scattering
experiments at room temperature. The measured spectra, from which
the scattering response at $B=\unit{0}{T}$ has been subtracted,
are shown in the form of an intensity false color map in
Fig.~\ref{fig:CRCroomtemp}. Electronic Raman scattering far away
from the Raman G band is not clearly observed, which is why the
spectral range in Fig. \ref{fig:CRCroomtemp} was limited to show
the Raman G band, only. We qualitatively observe again the same
behavior as at $T=\unit{4.2}{K}$. As it was already reported in
the case of K-point carriers in bulk graphite~\cite{Kossacki2011},
avoided crossings caused by the interaction of the zone-center
E$_{2g}$-phonon with electronic excitations L$_{0,1}$ and
L$_{-1,2}$ are clearly observed at room temperature, while a trace
of the interaction with L$_{-1,1}$ is also seen.

\section{Summary and conclusion} \label{sec:6}

We have discussed the magneto-Raman scattering response of
graphene on natural graphite measured in two different
polarization configurations: co-circular and crossed-circular.
Scattering signals from phonon- as well as from electronic
excitations of different types are identified in the data obtained
in the two configurations, allowing to verify the different
selection rules of the related scattering processes. The most
pronounced electronic Raman scattering signals are measured in the
co-circular configurations and they are due to electronic
excitations among LLs of graphene which are symmetric with respect
to the LL index, i.e. L$_{-n,n}$. The integrated intensities of
these signals are found to scale $\propto B/n$. Furthermore, the
strongest two of these excitations show a clearly asymmetric line
shape that depends on the applied magnetic field strength $B$. In
the crossed-circular configuration, the dominant feature measured
is the G band which shows a series of avoided crossings as a
function of $B$. The L$_{0,1}^{-1,0}$ is the only electronic
excitation observed far away from the phonon feature in this
configuration. The E$_{2g}$-phonon in this graphene system
interacts with the different electronic excitations L$_{-n,n}$,
L$_{-n,n+1}^{-n-1,n}$ as well as with L$_{0,2}^{-2,0}$. We
quantify the interaction strengths associated with the different
excitations and additionally extract the next-nearest-neighbor
hopping energy $\gamma_0'$ that describes the electron-hole
asymmetry in graphene. The unexpected observation, namely the G
band avoided crossings caused by other than L$_{-n,n+1}^{-n-1,n}$
excitations as well as the Raman scattering signals in the
co-circular configuration we assigned to usually infrared-active
electronic modes, are both possibly related to a residual
interaction between the graphene flake and the graphite substrate
on which it lies, and to the quasi backscattering configuration
used in this experiment. While the discussions in this report were
mainly focused on results obtained from measurements performed at
$\unit{4.2}{K}$, we show that most of these properties persist up
to room temperature. This holds especially as we readily observe
Raman scattering from electronic excitations as well as the
characteristic magnetic-field induced oscillations in the G band
feature, both at room temperature.

Our magneto-Raman scattering experiment presented here cannot
unambiguously determine the nature of these graphene-like
locations. The response from bulk graphite is also observed on
such locations giving an indication on the relatively small
thickness of the region of interest. Recent tight binding
calculations of the band structure of AA stacked
graphite~\cite{Lobato2011} show that this material host Dirac
fermions along the \textbf{HKH} corner of its Brillouin with Fermi
velocities differing by $3\%$, in line with the splitting of the
$L_{-n,n}$ excitations observed in this work. Nevertheless, the
value of the Fermi energy at the \textbf{H} and \textbf{K} points
are not compatible with the observation of any inter Landau level
excitations below $\sim\unit{800}{meV}$. The most probable
scenario is that the first graphene layer does not follow the
Bernal stacking sequence and is twisted with respect to the
underlying layer. The electronic properties of such twisted
graphene bilayers have recently been the subject of intense
theoretical research~\cite{Mele2011}, indicating not only a
renormalization of the Fermi velocity~\cite{Lopes2007} but the
existence of two distinct cones with different Fermi velocities.

To conclude, by using polarization-resolved magneto-Raman
scattering techniques, we have experimentally determined Raman
scattering selection rules for different inelastic light
scattering processes in graphene-like domains at the surface of
bulk graphite involving both phonon- and electronic excitations.
Interactions between both types of excitations have been
quantified as was the electron-hole asymmetry in our sample, which
could be directly measured and expressed in terms of the
next-nearest-neighbor hopping energy $\gamma_0'$. We furthermore
demonstrated that LL spectroscopy using magneto-Raman scattering
in graphene can be performed at room-temperature.

\section{Acknowledgements}
We warmly acknowledge fruitful discussions with D.M. Basko and
V.I. Fal'ko and technical support from Ivan Breslavetz. Part of
this work has been supported by GACR P204/10/1020, GRA/10/E006
(EPIGRAT), RTRA "DISPOGRAPH" projects and by EuroMagNET II under
the EU contract number 228043. P.K. is financially supported by
the EU under FP7, contract no. 221515 ''MOCNA''. Yu. L. is
supported by the Russian state contract No. 16.740.11.0146.


%

\end{document}